\documentclass[cits]{PoS}
\usepackage{amssymb,latexsym,amsmath}
\newcommand{\eq}[1]{\begin{align} #1 \end{align}}

\title{Fluctuations in Statistical Models}

\ShortTitle{Fluctuations in Statistical Models}

\author{\speaker{Mark Gorenstein}\\
\\
 Bogolyubov Institute for Theoretical Physics, Kiev, Ukraine.
        \\ E-mail: \email{mark@mgor.kiev.ua}}

\abstract{The multiplicity fluctuations of hadrons are studied
within the statistical hadron-resonance gas model in the large
volume limit. The role of quantum statistics and resonance decay
effects are discussed. The microscopic correlator method is used
to enforce conservation of three charges -- baryon number,
electric charge, and strangeness -- in the canonical ensemble. In
addition, in the micro-canonical ensemble energy conservation is
included. An analytical method is used to account for resonance
decays. The multiplicity distributions and the scaled variances
for negatively  and positively charged hadrons are calculated for
the sets of thermodynamical parameters along the chemical
freeze-out line of central Pb+Pb (Au+Au) collisions from SIS to
LHC energies. Predictions obtained within different statistical
ensembles are compared with the preliminary NA49 experimental
results on central Pb+Pb collisions in the SPS energy range. The
measured fluctuations are significantly narrower than the Poisson
ones and clearly favor expectations for  the micro-canonical
ensemble. Thus, this is a first  observation of the recently
predicted suppression of the multiplicity fluctuations in
relativistic gases in the thermodynamical limit due to
conservation laws. }

\FullConference{Critical Point and Onset of Deconfinement - 4th International Workshop\\
         July 9-13, 2007\\ Darmstadt, Germany}

\begin{document}

%\maketitle

%-----------------------------------------------------------------------
\section{Introduction}

%\vspace{-0.3cm}
\noindent
 For more than 50 years statistical
models of strong interactions \cite{fermi,landau,hagedorn} have
served as an important tool to investigate high energy nuclear
collisions. The main subject of the past study has been the mean
multiplicity of produced hadrons (see e.g. Refs.
\cite{stat1,FOC,FOP,pbm}). Only recently, due to a rapid
development of experimental techniques, first measurements of
fluctuations of particle multiplicity \cite{fluc-mult} and
transverse momenta \cite{fluc-pT} were performed. The growing
interest in the study of fluctuations in strong interactions (see
e.g., reviews \cite{fluc1}) is motivated by expectations of
anomalies in the vicinity of the onset of deconfinement \cite{ood}
and in the case when the expanding system goes through the
transition line between the quark-gluon plasma and the hadron gas
\cite{fluc2}. In particular, a critical point of strongly
interacting matter may be signaled by a characteristic power-law
pattern in fluctuations \cite{fluc3}.

There is a qualitative difference in the properties of the mean
multiplicity and the scaled variance of multiplicity distribution
in statistical models. In the case of the mean multiplicity
results obtained with the grand canonical ensemble (GCE),
canonical ensemble (CE), and micro-canonical ensemble (MCE)
approach  each other in the large volume limit. One refers here to
the thermodynamical equivalence of the statistical ensembles. It
was recently found  \cite{CE,res} that corresponding results for
the scaled variance are different in different ensembles, and thus
the scaled variance is sensitive to conservation laws obeyed by a
statistical system. The differences are preserved  in the
thermodynamic limit.

We calculate the multiplicity fluctuations in central collisions
of heavy nuclei within the MCE formulation of the hadron-resonance
gas model \cite{MCE}. Fluctuations are quantified by the ratio of
the variance of the multiplicity distribution and its mean value,
the so-called scaled variance. The model calculations are compared
with the corresponding preliminary results \cite{NA49} of NA49 on
central Pb+Pb collisions at the CERN SPS energies.

%-----------------------------------------------------------------------
\section{Statistical Models}

%\vspace{-0.3cm}
\noindent
 The mean multiplicities of positively, negatively
and all charged particles are defined as:
 \eq{
 \langle N_-\rangle \;=\; \sum_{i,q_i<0} \langle N_i\rangle\;,~~~~
 \langle N_{+}\rangle \;=\; \sum_{i,q_i>0} \langle
 N_i\rangle\;,~~~~
 \langle N_{ch}\rangle \;=\; \sum_{i,q_i\neq 0} \langle
 N_i\rangle\;,
 \label{pminch}
 }
where the average final state (after
resonance decays) multiplicities $\langle N_i\rangle$ are equal
to:
 \eq{\label{<N>}
 \langle N_i\rangle
 \;=\;
 \langle N_i^*\rangle + \sum_R \langle N_R\rangle \langle
 n_{i}\rangle_R\;.
 }
In Eq.~(\ref{<N>}), $N_i^*$ denotes the number of stable primary
hadrons of species $i$, the summation $\sum_R$ runs over all types
of resonances $R$,
and $\langle n_i\rangle_R \equiv \sum_r
b_r^R n_{i,r}^R$~ is the average  over resonance decay channels.
The parameters $b^R_r$ are the branching ratios of the $r$-th
branches, $n_{i,r}^R$ is the number of particles of species $i$
produced in resonance $R$ decays via a decay mode $r$. The index $r$
runs over all decay channels of a resonance $R$, with the
requirement $\sum_{r} b_r^R=1$.
In the GCE formulation of the hadron-resonance gas model
the mean number of stable
primary particles, $\langle N_i^* \rangle$, and the mean number of
resonances, $\langle N_R \rangle$, can be calculated as:
 \eq{\label{Ni-gce}
 \langle N_j\rangle \;\equiv\; \sum_{\bf p} \langle n_{{\bf p},j}\rangle
  \;=\; \frac{g_j V}{2\pi^{2}}\int_{0}^{\infty}p^{2}dp\; \langle
 n_{{\bf p},j}\rangle\;,
}
where $V$ is the system volume and $g_j$ is the degeneracy factor
of particle of the species $j$  (number of spin states). In the
thermodynamic limit, $V\rightarrow \infty$, the sum over the
momentum states can be substituted by a momentum integral.  The
$\langle n_{{\bf p},j} \rangle$ denotes the mean occupation number
of a single quantum state labelled by the  momentum vector ${\bf
p}$~,
 \eq{
 \langle n_{{\bf p},j} \rangle
 ~& = ~\frac {1} {\exp \left[\left( \epsilon_{{\bf p}j} - \mu_j \right)/ T\right]
 ~-~ \alpha_j}~, \label{np-aver}
  }
where $T$ is the system temperature, $m_j$ is the mass
of a particle $j$, $\epsilon_{{\bf p}j}=\sqrt{{\bf p}^{2}+m_j^{2}}$
is a single  particle energy. A value of $\alpha_j$ depends on quantum
statistics, it is $+1$ for bosons and $-1$ for fermions, while
$\alpha_j=0$ gives the Boltzmann approximation. The chemical
potential $\mu_j$ of a species $j$ equals to:
\eq{ \mu_j~=~q_j~\mu_Q~+~b_j~\mu_B~+~s_j~\mu_S ~,\label{chempot}}
where $q_j,~b_j,~s_j$ are the particle electric charge, baryon
number, and strangeness, respectively, while $\mu_Q,~\mu_B,~\mu_S$
are the corresponding chemical potentials which regulate the
average values of these global conserved charges in the GCE. In
the limit $V\rightarrow\infty$~, Eq.~(\ref{Ni-gce}-\ref{chempot})
are also valid for the CE and MCE, if the energy density and
conserved charge densities are the same in all three ensembles.
This is usually referred to as the thermodynamical equivalence of
all statistical ensembles. However, the thermodynamical
equivalence does not apply to fluctuations.

In statistical models a natural measure of multiplicity
fluctuations is the scaled variance of the multiplicity
distribution.
For negatively, positively, and all charged particles the
scaled variances read:
 \eq{
 \omega^- ~=~ \frac{\langle \left( \Delta N_- \right)^2
\rangle}{\langle N_-
  \rangle}~,~~~~
 \omega^+~ =~ \frac{\langle \left( \Delta N_+ \right)^2
\rangle}{\langle N_+
  \rangle}~,~~~~
 \omega^{ch}~ =~ \frac{\langle \left( \Delta N_{ch} \right)^2
\rangle}{\langle N_{ch}
  \rangle}~.
\label{omega-all}
 }
The variances in Eq.~(\ref{omega-all}) can be presented as a sum
of the correlators:
\eq{
\langle \left( \Delta N_- \right)^2 \rangle
 ~& =~
\sum_{i,j;~q_i<0,q_j<0} \langle \Delta N_i \Delta N_j
\rangle~,~~~~ \langle \left( \Delta N_+ \right)^2 \rangle
 ~ =~
\sum_{i,j;~q_i>0,q_j>0} \langle \Delta N_i \Delta N_j \rangle~,\nonumber \\
\langle \left( \Delta N_{ch} \right)^2 \rangle
 ~ &=~
\sum_{i,j;~q_i\neq 0,q_j\neq 0} \langle \Delta N_i \Delta N_j
\rangle~, \label{DNpm}
}
where $\Delta N_i\equiv N_i -\langle N_i\rangle$. The correlators
in Eq.~(\ref{DNpm}) include both the correlations between
primordial hadrons and those of final state hadrons due to the
resonance decays (resonance decays obey charge as well as energy-momentum conservation).

In the MCE, the energy and conserved charges are fixed exactly for
each microscopic state of the system.  The primordial (before
resonance decays) microscopic correlators in the MCE has the form
\cite{MCE}:
 \eq{\label{corr}
 &
 \langle \Delta n_{{\bf p},j}  \Delta n_{{\bf k},i} \rangle_{m.c.e.}
 ~=\;  \upsilon_{{\bf p},j}^2\,\delta_{ij}\,\delta_{{\bf p}{\bf k}}
 \;-\;  \frac{\upsilon_{{\bf p},j}^2v_{{\bf k},i}^2}{|A|}\;
 [\;q_iq_j M_{qq} + b_ib_j M_{bb} + s_is_j M_{ss} \nonumber
 \\
 &+ ~\left(q_is_j + q_js_i\right) M_{qs}~
 - ~\left(q_ib_j + q_jb_i\right) M_{qb}~
 - ~\left(b_is_j + b_js_i\right) M_{bs}\nonumber
 \\
 &+~ \epsilon_{{\bf p}j}\epsilon_{{\bf k}i} M_{\epsilon\epsilon}~-~
 \left(q_i \epsilon_{{\bf p}j} + q_j\epsilon_{{\bf k}i} \right)
 M_{q\epsilon}~
  +~ \left(b_i \epsilon_{{\bf p}j} + b_j\epsilon_{{\bf k}i} \right)
  M_{b\epsilon}~
  - ~\left(s_i \epsilon_{{\bf p}j} + s_j\epsilon_{{\bf k}i} \right) M_{s\epsilon}
 \;]\;,
 }
where $|A|$ is the determinant and $M_{ij}$ are the minors of the
following matrix,
 \eq{\label{matrix}
 A =
 \begin{pmatrix}
 \Delta (q^2) & \Delta (bq) & \Delta (sq) & \Delta (\epsilon q)\\
 \Delta (q b) & \Delta (b^2) & \Delta (sb) & \Delta (\epsilon b)\\
 \Delta (q s) & \Delta (b s) & \Delta (s^2) & \Delta (\epsilon s)\\
 \Delta (q \epsilon) & \Delta (b \epsilon) & \Delta (s \epsilon) & \Delta (\epsilon^2)
 \end{pmatrix}\;,
 }
with the elements, $\;\Delta (q^2)\equiv\sum_{{\bf p},j}
q_{j}^2\upsilon_{{\bf p},j}^2\;$, $\;\Delta (qb)\equiv \sum_{{\bf
p},j} q_{j}b_{j}\upsilon_{{\bf p},j}^2\;$, $\;\Delta
(q\epsilon)\equiv \sum_{{\bf p},j} q_{j}\epsilon_{{\bf
p}j}\upsilon_{{\bf p},j}^2\;$, etc. The sum, $\sum_{{\bf p},j}$~,
means integration over momentum ${\bf p}$, and the summation over
all hadron-resonance species~$j$ contained in the model. The first
term in the r.h.s. of Eq.~(\ref{corr}) corresponds to the
microscopic correlator  in the GCE. Note that a presence of the
terms containing a single particle energy, $\epsilon_{{\bf
p}j}=\sqrt{{\bf p}^{2}+m_j^{2}}$, in Eq.~(\ref{corr}) is a
consequence of  energy conservation. In the CE, only charges are
conserved, thus the terms containing $\epsilon_{{\bf p}j}$ in
Eq.~(\ref{corr}) are absent. The  $A$ in Eq.~(\ref{matrix})
becomes then the $3\times 3$ matrix (see Ref.~\cite{res}).  An
important property of the microscopic correlator method is that
the particle number fluctuations and the correlations in the MCE
or CE, although being different from those in the GCE, are
expressed by quantities calculated within the GCE. The microscopic
correlator (\ref{corr}) can be used  to calculate the primordial
particle correlator  in the MCE (or in the CE):
\eq{
  \langle \Delta N_{i} ~\Delta N_{j}~\rangle_{m.c.e.}
 &~= \sum_{{\bf p},{\bf k}}~\langle \Delta n_{{\bf p},i}~\Delta
 n_{{\bf k},j}\rangle_{m.c.e.}\;. \label{mc-corr-mce}
}

A second  feature of the MCE (or CE) is the modification of the
resonance decay contribution to the fluctuations in comparison to
the GCE results of Ref.~\cite{Koch}. In the MCE it reads
\cite{MCE}:
 \eq{
 \langle \Delta N_i\,\Delta N_j\rangle_{m.c.e.}
 ~&=\; \langle\Delta N_i^* \Delta N_j^*\rangle_{m.c.e.}
  \;+\; \sum_R \langle N_R\rangle\; \langle \Delta n_{i}\; \Delta  n_{j}\rangle_R
 \;+\; \sum_R \langle\Delta N_i^*\; \Delta N_R\rangle_{m.c.e.}\; \langle n_{j}\rangle_R
  \; \nonumber
 \\
 &+\; \sum_R  \langle\Delta N_j^*\;\Delta N_R\rangle_{m.c.e.}\; \langle n_{i}\rangle_R
  \;+\; \sum_{R, R'} \langle\Delta N_R\;\Delta N_{R'}\rangle_{m.c.e.}
  \; \langle n_{i}\rangle_R\;
       \langle n_{j}\rangle_{R^{'}}\;.\label{corr-MCE}
 }
Additional terms in Eq.~(\ref{corr-MCE}) compared to the GCE
results \cite{Koch} are due to the correlations (for primordial
particles) induced by energy and charge conservations in the MCE.
The Eq.~(\ref{corr-MCE}) has the same form in the CE \cite{res}
and MCE \cite{MCE}, the difference between these two ensembles
appears because of different microscopic correlators (\ref{corr}).
The microscopic correlators of the MCE together with
Eq.~(\ref{mc-corr-mce}) should be used to calculate
  the correlators $\langle\Delta N_i^*
\Delta N_j^*\rangle_{m.c.e.}$~,~$\langle\Delta N_i^*\; \Delta
N_R\rangle_{m.c.e.}~$, $~\langle\Delta N_j^*\;\Delta
N_R\rangle_{m.c.e.}~$, $~\langle\Delta N_R\;\Delta
N_{R'}\rangle_{m.c.e.}$ entering in Eq.~(\ref{corr-MCE}) .

The microscopic correlators and the scaled variance are connected
with the width of the multiplicity distribution. It has been shown
in Ref.~\cite{CLT} that in statistical models the form of the
multiplicity distribution derived within any ensemble (e.g. GCE,
CE and MCE) approaches the Gauss distribution:
\begin{equation}\label{Gauss}
P_G(N) = \frac{1}{\sqrt{2 \pi ~\omega~\langle N \rangle}} ~\exp \left[ -
\frac{\left(N~-~\langle N \rangle \right)^2}{2 ~\omega ~\langle N \rangle} \right]~
\end{equation}
in the large volume limit i.e. $\langle N \rangle \rightarrow \infty$.
The width of this Gaussian, $\sigma = \sqrt{\omega~ \langle N \rangle}$, is determined by
the choice of the statistical ensemble,  while from the
thermodynamic equivalence of the statistical ensembles
it follows that the expectation value $\langle N \rangle$ remains the same.

\section{Multiplicity Fluctuations at the Chemical Freeze-out }\label{sec-HG}

%\vspace{-0.3cm}
\noindent
 Once a suitable set of thermodynamical parameters,
$T,\mu_B,\gamma_S$, is determined in central nucleus-nucleus
collisions for each collision energy, the scaled variance of
negatively, positively, and all charged particles can be
calculated using Eqs.~(\ref{omega-all}-\ref{DNpm}).  The
$\omega^{-}$ and $\omega^+$ in different statistical ensembles are
presented in Fig.~1  for different collision energies. The values
of $\sqrt{s_{NN}}$ marked in Fig.~1 correspond to the beam
energies at SIS (2$A$~GeV), AGS (11.6$A$~GeV), SPS ($20A$, $30A$,
$40A$, $80A$, and $158A$~GeV), colliding energies at RHIC
($\sqrt{s_{NN}}=62.4$~GeV, $130$~GeV, and $200$~GeV), and LHC
($\sqrt{s_{NN}}=5500$~GeV). The mean multiplicities,  $\langle
N_i\rangle$, used for calculation of the scaled variance (see
Eq.~(\ref{omega-all})) are given by Eqs.~(\ref{<N>}) and
(\ref{Ni-gce}) and remain the same in all three ensembles. The
variances in Eq.~(\ref{omega-all}) are calculated using the
corresponding correlators $\langle \Delta N_i \Delta N_j \rangle$
in the GCE, CE, and MCE. For the calculations of final state
correlators the summation in Eq.~(\ref{corr-MCE}) should include
all resonances $R$ and $R^{\prime}$ which have particles of the
species $i$ and/or $j$ in their decay channels.
\begin{figure}[ht!]
\begin{center}
 \includegraphics[scale=0.53]{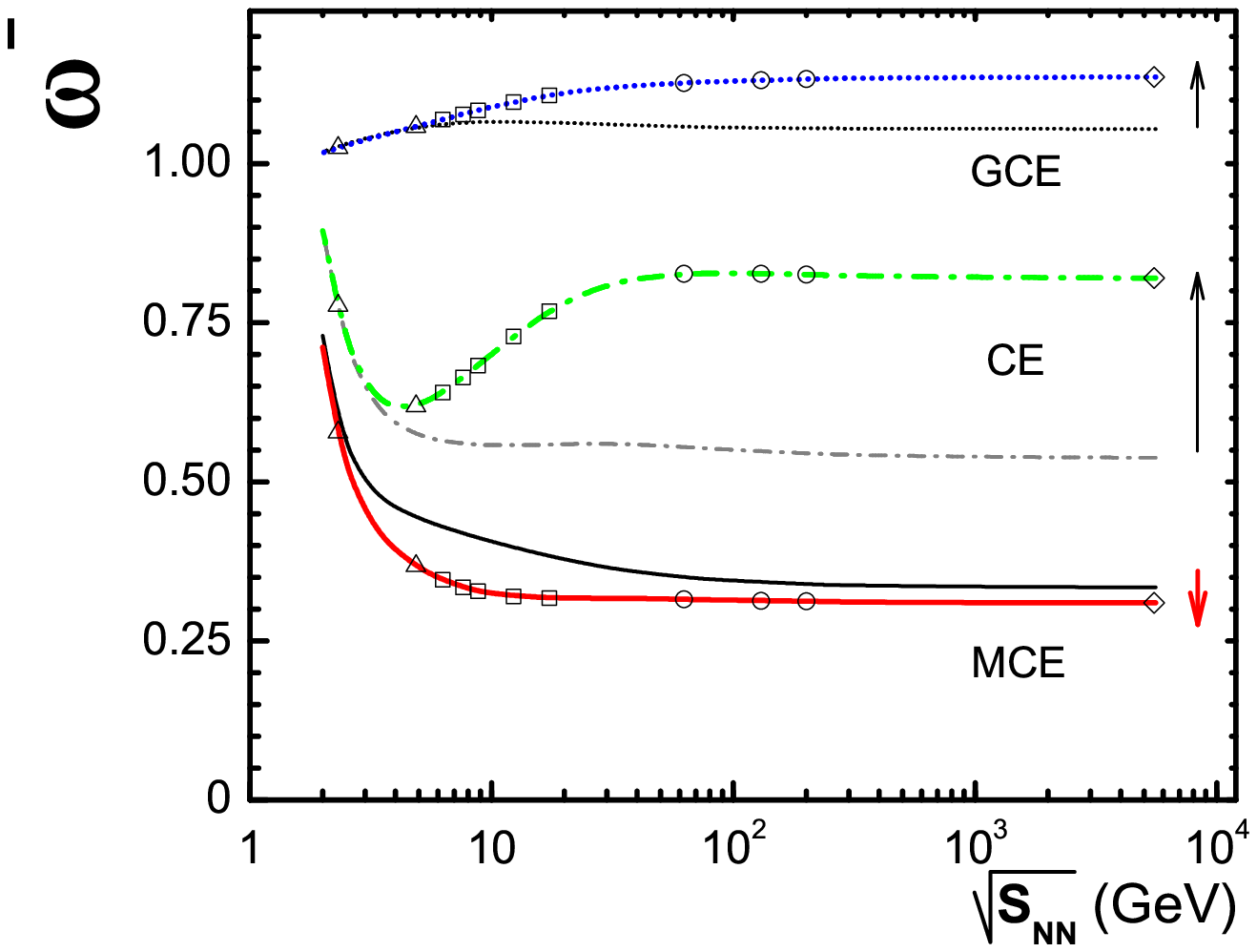}
 \includegraphics[scale=0.53]{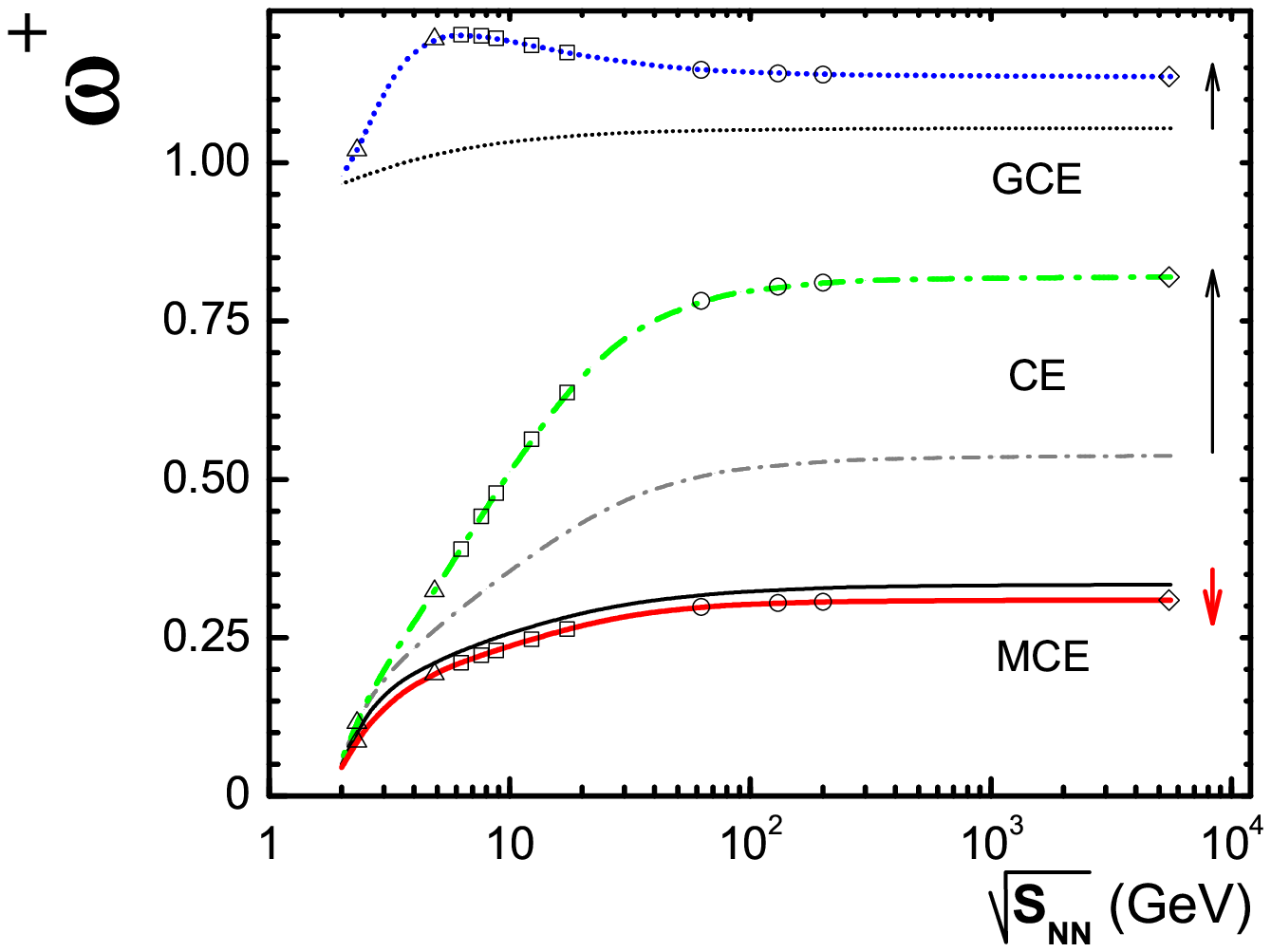}
 \caption{The scaled variances for negatively $\omega^-$ (left)
and positively $\omega^+$ (right) charged particles. Both
primordial and final $\omega^{\pm}$ are shown along the chemical
freeze-out line for central Pb+Pb (Au+Au) collisions as the
functions of the c.m. energy of the nucleon pair $\sqrt{s_{NN}}$.
Different lines present the GCE, CE, and MCE results. Symbols at
the lines for final particles correspond to the specific
accelerator collision energies. The arrows show the effect of
resonance decays.} \label{omega_m}
\end{center}
\end{figure}

%\begin{figure}[ht!]
%\begin{center}
%\includegraphics[scale=0.9]{fig3.eps}
% \caption{The same as in Figs.~\ref{omega_m}, but for $\omega^{ch}$.}
% \label{omega_ch}
%\end{center}
%\end{figure}
At the chemical freeze-out of heavy-ion collisions, the Bose
effect for pions and resonance decays are important  and thus (see
also Ref.~\cite{res}): $\omega^-_{g.c.e.}\cong 1.1$ and
$\omega^+_{g.c.e.}\cong 1.2$ at the SPS energies. Note that in the
Boltzmann approximation and neglecting the resonance decay effect
one finds $\omega^-_{g.c.e.}=\omega^+_{g.c.e.}=1$.

Some qualitative features of the results should be mentioned. The
effect of Bose and Fermi statistics is seen in primordial values
in the GCE. At low temperatures most of positively charged hadrons
are protons, and Fermi statistics dominates,
$\omega^{+}_{g.c.e.}<1$. On the other hand, in the limit of high
temperature (low $\mu_B/T$) most charged hadrons are pions and the
effect of Bose statistics dominates, $\omega_{g.c.e.}^{\pm}>1$.
Along the chemical freeze-out line, $\omega_{g.c.e.}^-$ is always
slightly larger than 1, as $\pi^-$ mesons dominate at both low and
high temperatures. The bump in $\omega^+_{g.c.e.}$ for final state
particles seen in Fig.~1 at the small collision energies  is due
to a correlated production of proton and $\pi^+$ meson from
$\Delta^{++}$ decays. This single resonance contribution dominates
in $\omega^+_{g.c.e.}$ at small collision energies (small
temperatures), but becomes relatively unimportant at the high
collision energies.

A minimum in $\omega_{c.e.}^{-}$ for final particles is seen in
Fig.~\ref{omega_m}. This is due to two effects. As the number of
negatively charged particles is relatively small, $\langle
N_-\rangle \ll \langle N_+\rangle$, at the low collision energies,
both the CE suppression and the resonance decay effect are small.
With increasing $\sqrt{s_{NN}}$, the CE effect alone leads to a
decrease of $\omega^-_{c.e}$, but the resonance decay effect only
leads to an increase of $\omega^-_{c.e}$. A combination of these
two effects, the CE suppression and the resonance enhancement,
leads to a minimum of $\omega^-_{c.e}$.

As expected, $\omega_{m.c.e.}<\omega_{c.e.}$, as an energy
conservation further suppresses the particle number fluctuations.
A new unexpected feature of the MCE  is the suppression of the
fluctuations after resonance decays. This is discussed in details
in Ref.~\cite{MCE}.

\section{Comparison with NA49 Data}

%\vspace{-0.3cm}
\noindent The scaled variance for the accepted
particles is assumed to be equal to (see discussion of this point
in Ref. \cite{MCE}):
\begin{align}\label{ac4}
\omega~\equiv~\frac{\langle n^2 \rangle~-~\langle n
\rangle ^2}{\langle n\rangle}~  =~1~-~q~ +q\cdot \omega_{4\pi}~,
\end{align}
where $\omega_{4\pi}$ is the scaled variance for the full
$4\pi$-acceptance.  In the large acceptance limit ($q \approx 1$)
the distribution of measured particles approaches the distribution
in the full acceptance. For a very small acceptance ($q \approx
0$) the measured distribution approaches the Poisson one
independent of the shape of the distribution in the full
acceptance.

% \subsection{Comparison with the NA49 Data for $\omega^-$ and $\omega^+$}

%
\begin{figure}[h!]
\includegraphics[scale=0.54]{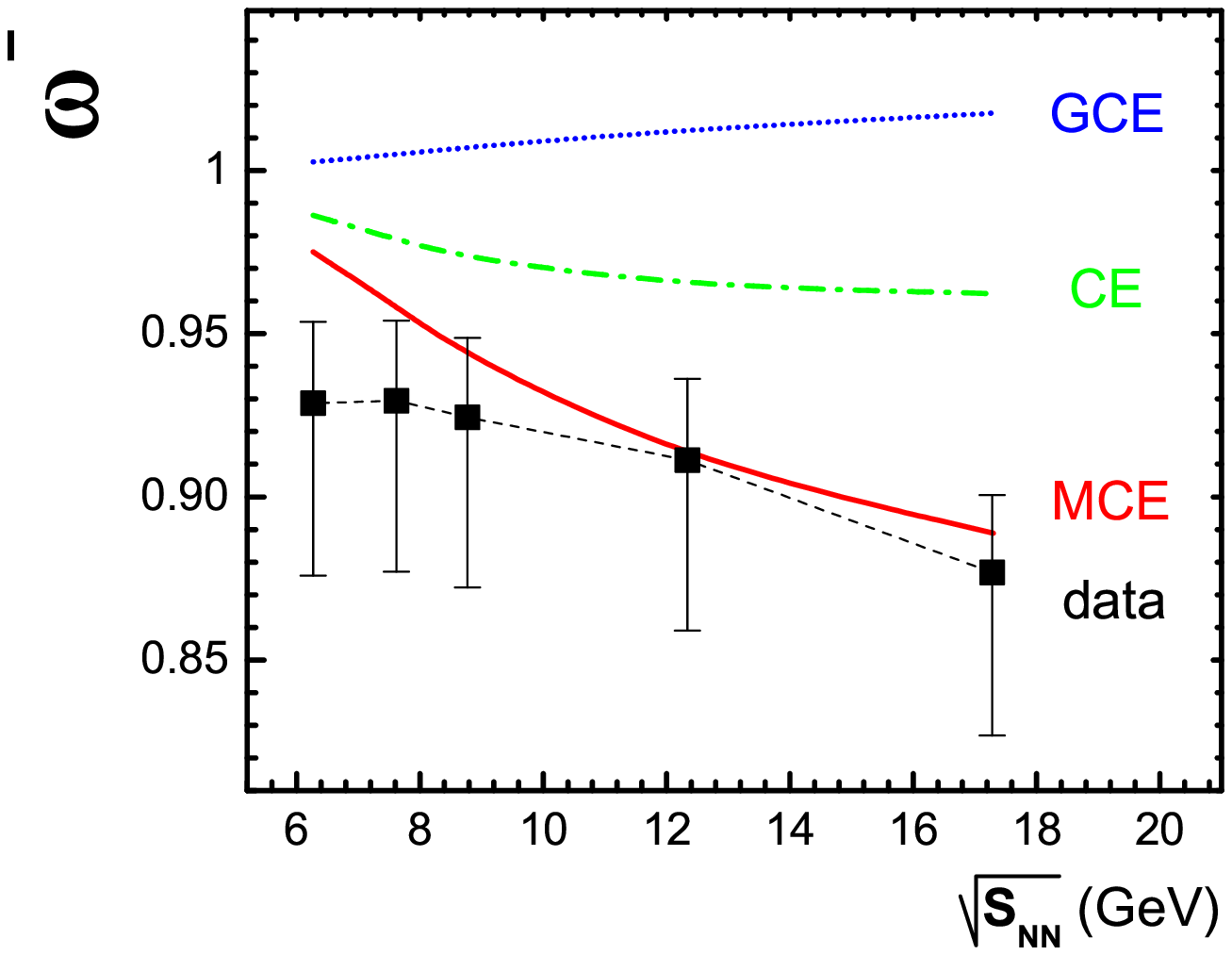}
\includegraphics[scale=0.54]{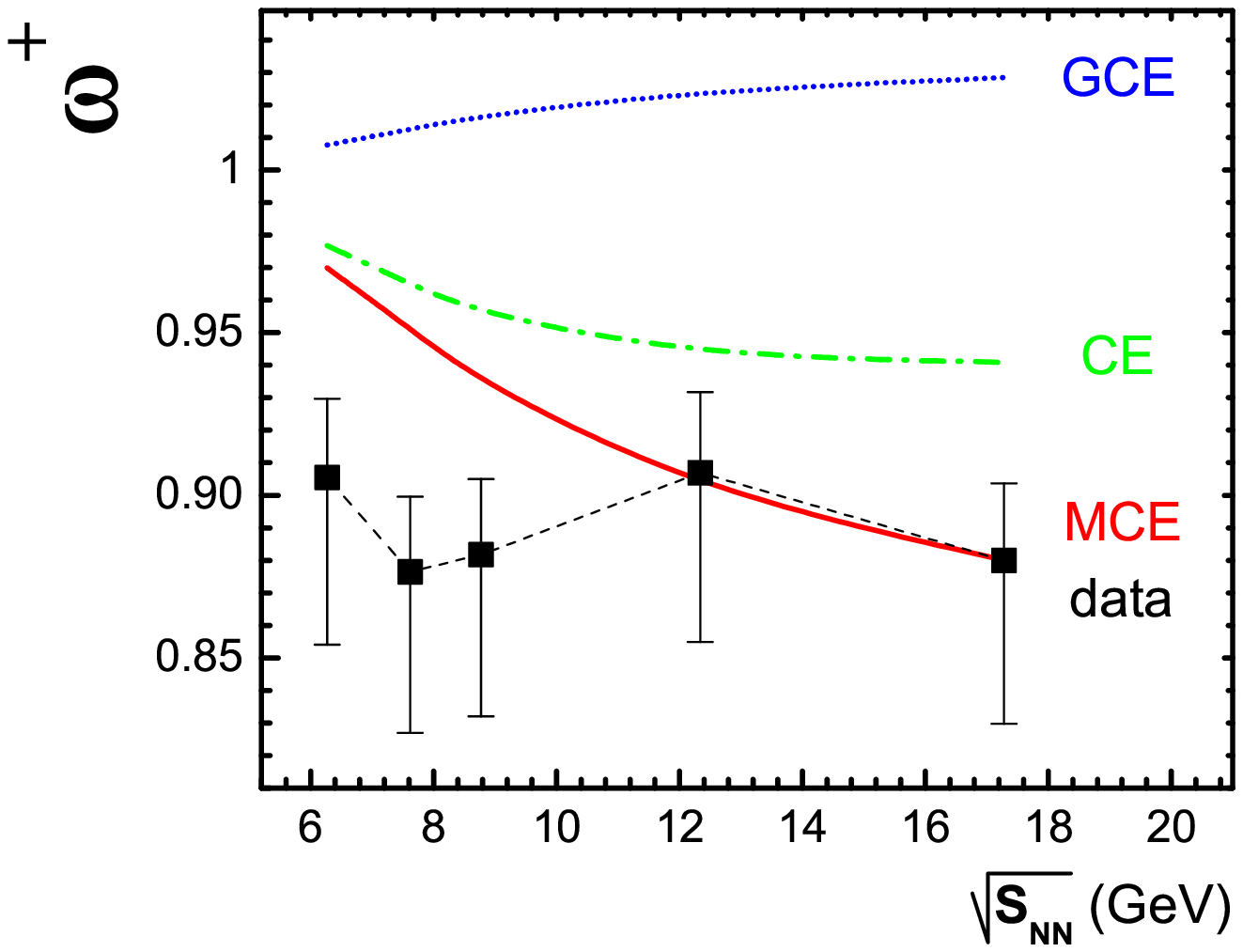}
 \caption{The scaled variances for negative (left)
and positive (right) hadrons along the chemical freeze-out line
for central Pb+Pb collisions at the SPS energies. The points show
the preliminary data of NA49 [17]. Total (statistical+systematic)
errors are indicated. The statistical model parameters $T$,
$\mu_B$, and $\gamma_S$ at different SPS collision energies are
taken at the chemical freeze-out, i.e. from fitting the hadron
yields. Lines show the GCE, CE, and MCE results. The NA49
experimental acceptance is approximately taken into account
according to Eq.~(4.1).}\label{stat-acc}
\end{figure}
The Fig.~\ref{stat-acc} presents the scaled variances $\omega^-$
and $\omega^+$ calculated with Eq.~(\ref{ac4}). The
hadron-resonance gas calculations in the GCE, CE, and MCE shown in
Fig.~1 are used for the $\omega_{4\pi}^{\pm}$. The NA49 acceptance
used for the fluctuation measurements is located in the forward
hemisphere \cite{NA49}). The acceptance probabilities for
positively and negatively charged hadrons are approximately equal,
$q^+\approx q^-$, and the numerical values at different SPS
energies are: $q^{\pm}=0.038, ~0.063,$ $0.085, ~0.131, ~0.163$ at
$\sqrt{s_{NN}}=6.27,~7.62,~8.77,~12.3,~17.3$~GeV, respectively.
Eq.~(\ref{ac4}) has the following property: if $\omega_{4\pi}$ is
smaller or larger than 1, the same inequality remains to be valid
for $\omega$ at any value of $0<q\le 1$. Thus one has a strong
qualitative difference between the predictions of the statistical
model valid for any freeze-out conditions and experimental
acceptances. The CE and MCE correspond to
$\omega_{m.c.e.}^{\pm}<\omega^{\pm}_{c.e.}<1$, and the GCE to
$\omega_{g.c.e.}^{\pm}> 1$.

From Fig.~\ref{stat-acc} it follows that the NA49 data for
$\omega^{\pm}$ extracted from 1\% of the most central  Pb+Pb
collisions at all SPS energies are best described by the results
of the hadron-resonance gas model calculated within the MCE. The
data reveal even stronger suppression of the particle number
fluctuations. The chemical freeze-out parameters found at fixed
collision energy have some uncertainties. However, the scaled
variances $\omega_{m.c.e.}^-$ and $\omega_{m.c.e.}^+$ calculated
in the full phase space within the MCE vary by less than 1\% when
changing the parameter set. In the NA49 acceptance the difference
is almost completely washed out.

%\subsection{Comparison of Distributions}

\begin{figure}[h!]
\label{DistNegPlot}
\includegraphics[scale=0.7]{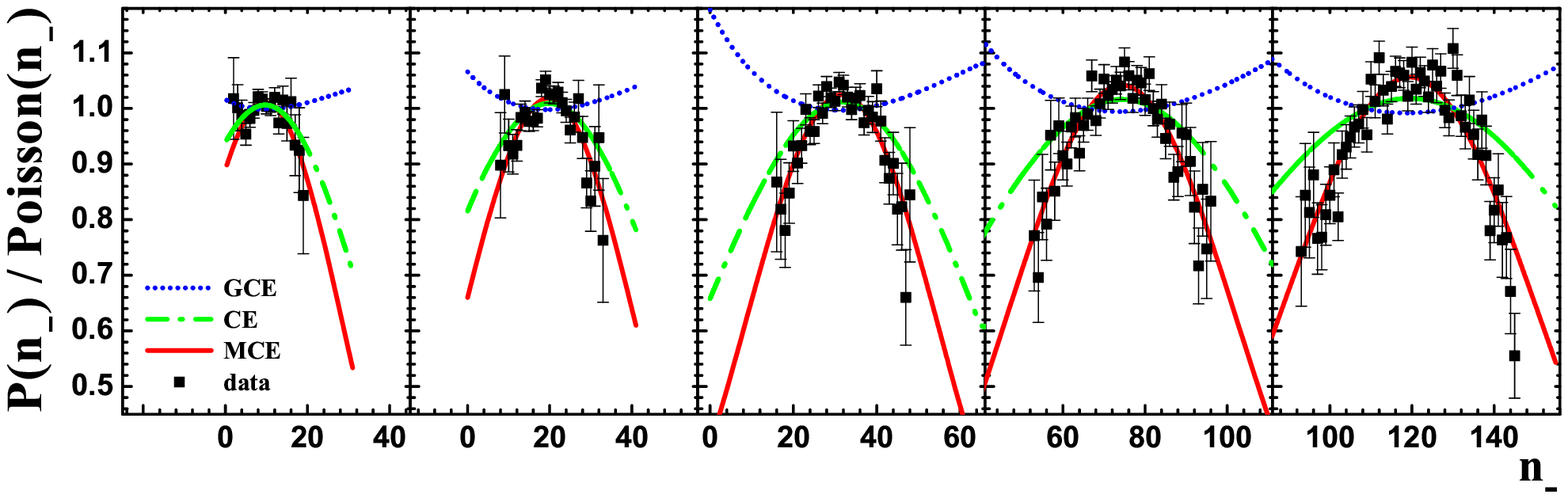}
\includegraphics[scale=0.7]{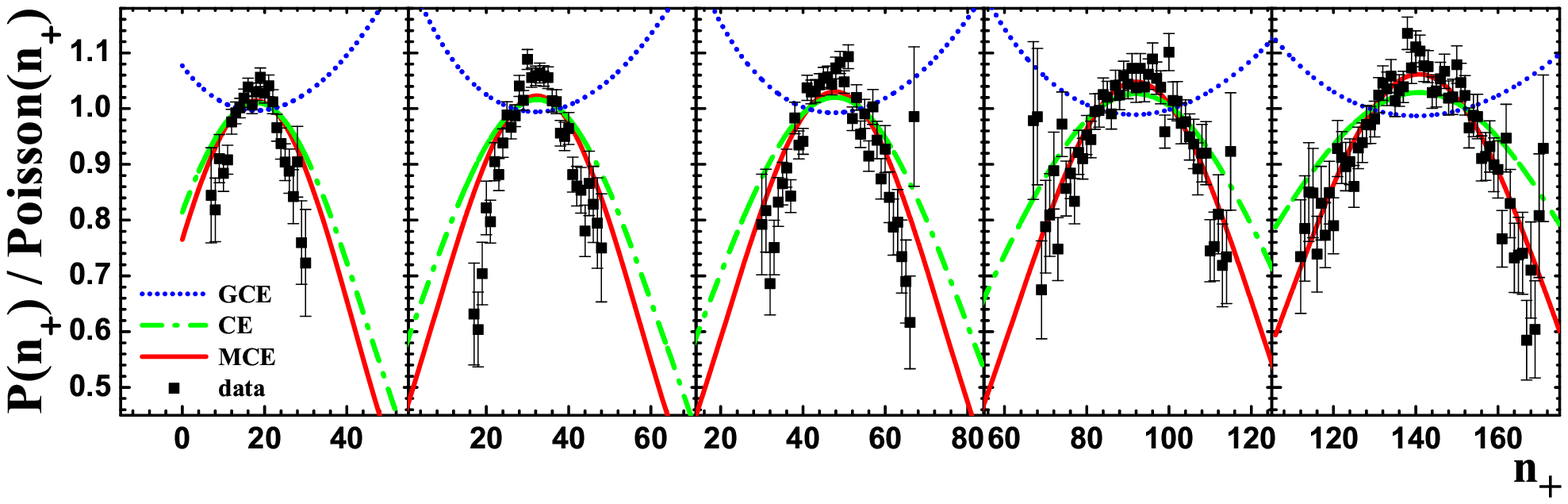}
\caption{
The ratio of the multiplicity distributions to Poisson ones for negatively
(upper panel) and positively (lower panel)
charged hadrons produced in central (1\%) Pb+Pb collisions
at 20$A$~GeV, 30$A$~GeV, 40$A$~GeV, 80$A$~GeV, and 158$A$~GeV
(from left to right)
in the NA49 acceptance \protect\cite{NA49}.
The preliminary experimental data (solid points)
of NA49 \protect\cite{NA49} are
compared with the prediction of the hadron-resonance gas
model obtained within different statistical ensembles,
the GCE (dotted lines), the CE (dashed-dotted lines), and
the MCE (solid lines).
}
\label{data2}
\end{figure}

In order to allow for a detailed comparison of the distributions
the ratio of the data and the model distributions to the Poisson
one is presented in Fig.~\ref{data2}.

The results for negatively and positively charged hadrons at
20$A$~GeV, 30$A$~GeV, 40$A$~GeV, 80$A$~GeV, and 158$A$~GeV are
shown in Fig.~3. The convex shape of the data reflects the fact
that the measured distribution is significantly narrower than the
Poisson one. This suppression of fluctuations is observed for both
charges, at all five SPS energies and it is consistent with the
results for the scaled variance shown and discussed previously.
The GCE hadron-resonance gas results are broader than the
corresponding Poisson distribution. The ratio has a concave shape.
An introduction of the quantum number conservation laws (the CE
results) leads to the convex shape and significantly improves
agreement with the data. Further improvement of the agreement is
obtained by the additional introduction of the energy conservation
law (the MCE results). The measured spectra surprisingly well
agree with the MCE predictions.

\section{Summary and Conclusions}

%\vspace{-0.3cm}
\noindent
The hadron multiplicity fluctuations in relativistic
nucleus-nucleus collisions have been predicted in the statistical
hadron-resonance gas model within the grand canonical, canonical,
and micro-canonical ensembles in the thermodynamical limit. The
microscopic correlator method has been extended to include three
conserved charges -- baryon number, electric charge, and
strangeness -- in the canonical ensemble, and additionally an
energy conservation in the micro-canonical ensemble. The
analytical formulas are used for the resonance decay contributions
to the correlations and fluctuations. The scaled variances of
negatively and  positively charged particles for primordial and
final state hadrons have been calculated at the chemical
freeze-out in central Pb+Pb (Au+Au) collisions for different
collision energies from SIS  to LHC.

The effect of Bose enhancement and Fermi suppression can be seen
in the primordial (before resonance decay) values of the scaled
variances.  The results presented in Fig. 1 demonstrate that the
effects of quantum statistics are small at the chemical
freeze-out. Resonance decays included into the GCE and CE lead to
the enhancement of particle number fluctuations. An important
feature of the MCE  is the suppression of the fluctuations after
resonance decays. This is discussed in details in Ref.~\cite{MCE}.

A comparison of the multiplicity distributions and the scaled
variances with the preliminary NA49 data on  Pb+Pb collisions at
the SPS energies has been done for the samples of about 1\% of
most central collisions selected by the number of projectile
participants. This selection allows to eliminate effect of
fluctuations of the number of nucleon participants. The effect of
the limited experimental acceptance was taken into account by use
of the uncorrelated particle approximation. The   measured
multiplicity distributions are significantly narrower than the
Poisson one and allow to distinguish between model results derived
within different statistical ensembles. The data surprisingly well
agree with the expectations for the micro-canonical ensemble and
exclude the canonical and grand-canonical ensembles. Thus, this is
a first experimental observation of the predicted suppression
\cite{CE,res,MCE} of the multiplicity fluctuations in relativistic
gases in the thermodynamical limit due to conservation laws.

A validity of the micro-canonical description is surprising. In
fact, significant event-by-event fluctuations of statistical model
parameters may be expected. For instance, only a part of the total
energy is available for the hadronization process. This part
should be used in the hadron-resonance gas calculations while the
remaining energy is contained in the collective motion of matter.
The ratio between the hadronization and collective energies may
vary from collision to collision and consequently increase the
multiplicity fluctuations. The agreement between the data and the
MCE predictions is even more surprising when the  processes which
are beyond the statistical hadron-resonance gas model are
considered. Examples of these are jet and mini-jet production,
heavy cluster formation, effects related to the phase transition
or instabilities of the quark-gluon plasma. Naively all of them
are expected to increase multiplicity fluctuations and thus lead
to a disagreement between the data and the MCE predictions. A
comparison of the data with the models which include these
processes is obviously needed for significant conclusions.

On the model side there are, however, at least 2 additional
effects which may lead to a suppression of the multiplicity
fluctuations. The first of them follows from improving the
description of the effect of the limited experimental acceptance
within MCE \cite{Hauer}. The second one follows from taking into
account the finite proper volume of hadrons. As shown  in
Ref.~\cite{ExclVol} the excluded volume effects lead to a
reduction of the particle number fluctuations. The quantitative
estimates of these two effects are needed.

More differential data on multiplicity fluctuations and
correlations are required for further tests of the validity of the
statistical models and observation of possible signals of the
phase transitions. The experimental resolution in a measurement of
the enhanced fluctuations due to the onset of deconfinement can be
increased by increasing acceptance.

\begin{acknowledgments}
I would like to thank   F. Becattini, V.V. Begun, M.~Ga\'zdzicki,
M. Hauer, A. Ker\"anen, A.P. Kostyuk, V.P.~Konchakovski,
B.~Lungwitz, and O.S.~Zozulya for fruitful collaborations. I am
also thankful to  E.L.~Bratkovskaya, A.I.~Bugrij, W.~Greiner,
I.N.~Mishustin, St.~Mr\'owczy\'nski, L.M.~Satarov, and
H.~St\"ocker, for numerous discussions.

\end{acknowledgments}

\end{document}